\documentclass[%
reprint,
superscriptaddress,
amsmath,amssymb,
aps,
]{revtex4-2}

\usepackage{graphicx}% Include figure files
\usepackage{dcolumn}% Align table columns on decimal point
\usepackage{bm}% bold math
\usepackage{mathrsfs}
\usepackage{float}
\usepackage{physics}
\usepackage{siunitx}
\usepackage[colorlinks,linkcolor=blue,anchorcolor=blue,urlcolor=blue,citecolor=blue]{hyperref}
\begin{document}

\preprint{APS/123-QED}

\title{Confined non-Hermitian skin effect in a semi-infinite Fock-state lattice}% Force line breaks with \\

\author{Zhi Jiao Deng}
\email{dengzhijiao926@hotmail.com}
\affiliation{College of Science, National University of Defense Technology, Changsha, Hunan 410073, China\\}
\affiliation{Hunan Key Laboratory of Mechanism and Technology of Quantum Information, Changsha, Hunan 410073, China\\} 
\affiliation{Key Laboratory of Low Dimensional Quantum Structures and Quantum Control of Ministry of Education, Hunan Normal University, Changsha, Hunan 410081, China\\}

\author{Xing Yao Mi}
\affiliation{College of Science, National University of Defense Technology, Changsha, Hunan 410073, China\\}

\author{Ruo Kun Cai}
\affiliation{College of Science, National University of Defense Technology, Changsha, Hunan 410073, China\\}

\author{Chun Wang Wu}
\email{cwwu@nudt.edu.cn}
\affiliation{College of Science, National University of Defense Technology, Changsha, Hunan 410073, China\\}
\affiliation{Hunan Key Laboratory of Mechanism and Technology of Quantum Information, Changsha, Hunan 410073, China\\}

\author{Ping Xing Chen}
\affiliation{College of Science, National University of Defense Technology, Changsha, Hunan 410073, China\\}
\affiliation{Hunan Key Laboratory of Mechanism and Technology of Quantum Information, Changsha, Hunan 410073, China\\}

\date{\today}% It is always \today, today,

\begin{abstract}
In this paper, we investigate the non-Hermitian skin effect in a semi-infinite Fock-state lattice, where the inherent coupling scales as $\sqrt{n}$. By analytically solving a non-uniform, non-reciprocal SSH model, we demonstrate that the intrinsic inhomogeneous coupling, in combination with non-reciprocity, fundamentally modifies the conventional skin effect. Instead of accumulating at the physical boundary, all eigenmodes become compressed and skewed within a finite spatial range determined by the inhomogeneous profile—a phenomenon we term the confined non-Hermitian skin effect. Consequently, the evolution of the probability distribution on the lattice starting from a single site is doubly confined: it is spatially bounded to a finite range by the inhomogeneous coupling, and further restricted to a one-sided trajectory at the edge of this range by the non-reciprocity. Moreover, a feasible experimental scheme based on a single trapped ion is also proposed. This work reveals how engineered coupling profiles in synthetic dimensions can reshape non-Hermitian properties and enable new protocols for quantum state manipulation. 
\end{abstract}

\maketitle

%\tableofcontents

\section{\label{sec:level1}Introduction}

In recent years, the topological physics of open systems has attracted a lot of attention \cite{zongshu1}. Due to the non-Hermitian nature of their Hamiltonians, the eigenvalues of such systems are generally complex, and the eigenstates no longer satisfy orthogonality \cite{zongshu2}. These features give rise to a variety of topological properties and phenomena \cite{zongshu1,fenlei,node1,node2,node3,node4} distinct from those in traditional Hermitian systems. One particularly important phenomenon is the non-Hermitian skin effect \cite{zongshu3,zongshu4,zongshu5}, where the bulk states become exponentially localized at the boundaries under open boundary conditions. This effect breaks the conventional bulk-boundary correspondence \cite{broken1,broken2} and has led to the development of new theoretical frameworks \cite{fangfa1,fangfa2,fangfa3,fangfa4,fangfa5,fangfa6}, notably the generalized Brillouin zone \cite{fangfa1}.

Building on this effect, researchers have uncovered many intriguing phenomena not present in Hermitian systems, such as self-acceleration \cite{acceleration}, self-healing \cite{healing}, edge burst \cite{burst}, chiral damping \cite{damping}, and reflected wave skin effects \cite{reflected}. It has also inspired proposals for unidirectional amplification \cite{amplification1,amplification2} and highly sensitive sensors \cite{sensor}. Experimentally, the characteristics of the non-Hermitian skin effect and its resulting dynamical behaviors have been demonstrated mainly in classical systems such as optics \cite{optics1,optics2,optics3}, electrical circuits \cite{electronics1,electronics2}, and mechanical setups \cite{mechanics1,mechanics2}. In contrast, its experimental realizations in quantum systems are rare. Owing to technical challenges, observation has so far been reported on cold-atom platform with only a few unit cells \cite{coldatom}. Although, this effect is theoretically robust and persists in the thermodynamic limit. 

On the other hand, the Fock-state lattice \cite{fock1,fock2} overcomes the limitation of finite lattice sites by adopting the concept of synthetic dimensions \cite{synthetic1,synthetic2}. It uses the excitation number $n$ as a lattice coordinate that is theoretically unbounded. Its natural compatibility with quantum optics platforms further facilitates the experimental observation of rich topological phenomena. However, a key feature of this lattice is the non-uniform coupling between adjacent sites, which introduces a position-dependent factor proportional to $\sqrt{n}$. This non-uniformity breaks translational symmetry and can lead to novel physical effects \cite{fock3,fock4}. At the same time, the semi-infinite nature of the Fock space complicates analytical treatment, yet a straightforward numerical diagonalization requires a finite cutoff, inevitably distorting part of the eigenstates. Previous studies have often either ignored this non-uniformity \cite{fock5,fock6} or restricted analyses to finite-dimensional subspaces with conserved excitation numbers \cite{fock3,fock4,fock7}. Recently, within the context of a Su-Schrieffer-Heeger (SSH) model constructed in a semi-infinite Fock-state lattice, this problem has been circumvented by obtaining analytical solutions for its energy spectrum and dynamics via a unitary transformation \cite{women}. This naturally raises a further question: can the semi-infinite Fock-state lattice simulate the non-Hermitian skin effect, and what distinctive features might arise from its inherent non-uniform coupling? While a preliminary proposal \cite{fock8} has suggested such a simulation, it did not account for the intrinsic non-uniformity in the spectral calculations. Thus, the core questions remain unanswered. 

This paper systematically addresses the above questions by studying a semi‑infinite, non‑reciprocal, and non‑uniformly coupled SSH model on a Fock‑state lattice. Using a similarity transformation, exact analytical solutions are obtained, fully characterizing the influence of non‑uniform coupling in the thermodynamic limit. These solutions uncover a novel skin characteristic: in contrast to the conventional skin effect where all bulk states are boundary-localized, here they shift towards the boundary within a finite range, with each state pinned around a specific lattice site. We term this phenomenon the confined non‑Hermitian skin effect. Correspondingly, its dynamical evolution also exhibits two key features: the inhomogeneous coupling confines the probability flow from a given site to a bounded spatial range, while the non-reciprocal coupling further retains only the probability distribution along the farthest feasible trajectory in the skin direction. This work clarifies how non‑uniform coupling modifies the skin effect and offers new possibilities for designing quantum state manipulation through engineered inhomogeneities in synthetic dimensions.

The paper is structured as follows. Section II details the construction of a semi-infinite, nonreciprocal SSH model on a Fock-state lattice and provides exact analytical solutions for its energy spectrum, eigenstates and dynamical evolution. Based on these solutions, Section III analyzes the eigenstate distributions, comparing them with the conventional non-Hermitian skin effect under uniform coupling and presenting the resulting dynamical features. Experimental feasibility with a single trapped ion is discussed in Section IV, and a summary is given in Section V.

\section{Model and Exact Solutions} 
\begin{figure}[t]
	\includegraphics{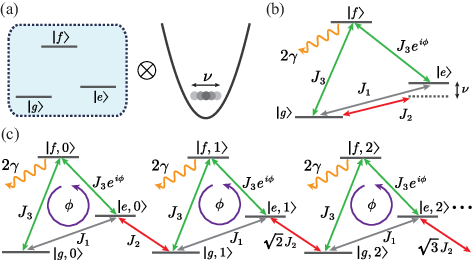}% Here is how to import EPS art
	\caption{\label{figure1} Schematic of the nonreciprocal, semi-infinite SSH model. (a) Decomposition of the system into two degrees of freedom: an internal three-level subsystem and external harmonic motion. (b) Coupling scheme among the energy levels. The coupling strengths $J_1$, $J_2$, and $J_3$ are all real. All couplings are resonant except for $J_2$, which is red-detuned by a phonon frequency $v$. The auxiliary level $|f\rangle$ decays to other levels (not shown) at a rate $2\gamma$. (c) Coupling structure between lattice sites, where each triangle $\{ |g,n\rangle,\ |e,n\rangle,\ |f,n\rangle \}$ represents a unit cell.}
\end{figure}

The system is a single trapped atom [Fig.~\ref{figure1}(a)] composed of three electronic levels $|g\rangle$, $|e\rangle$, $|f\rangle$ and an external harmonic oscillator mode of frequency $v$, with its state space spanned by the Fock states $\{ |n\rangle \}$. A semi-infinite SSH model is constructed on the combined manifold of the internal levels $|g\rangle$ and $|e\rangle$ and the external vibrational states $\{ |n\rangle \}$. This is achieved by coupling the levels $|g\rangle$ and $|e\rangle$ with two fields: a resonant field of strength $J_1$, and a field of strength $J_2$ that is red-detuned from resonance by one phonon frequency $v$ \cite{women}. Non-reciprocity is introduced via a dissipative coupling \cite{dissipative1,dissipative2}, which is realized by resonantly linking both $|g\rangle$ and $|e\rangle$ to an auxiliary level $|f\rangle$ that decays at a rate $2\gamma$ [Fig.~\ref{figure1}(b)]. Under the condition that no decay occurs, and setting $\hbar = 1$, the Hamiltonian of the system is given by \cite{Nojumps1,Nojumps2},
\begin{equation}
	\label{ham}
	\begin{split}
		\hat{H} = (&\, J_1 \left| e \right\rangle \left\langle g \right| + J_2 \hat{a} \left| e \right\rangle \left\langle g \right| + J_3 e^{-i\phi} \left| e \right\rangle \left\langle f \right| \\
		&+ J_3 \left| g \right\rangle \left\langle f \right| + \text{H.c.}) - i\gamma \left| f \right\rangle \left\langle f \right|,
	\end{split}
\end{equation}
where $\hat{a}$ and $\hat{a}^\dagger$ are the annihilation and creation operators for the vibrational mode. $J_3$ is the coupling strength from level $|g\rangle$ or $|e\rangle$ to level $|f\rangle$, with the latter transition carrying an additional phase $\phi$.

Figure~\ref{figure1}(c) shows the lattice coupling structure: each triangle represents a unit cell; specifically, the $n$-th triangle corresponds to the $n$-th cell, which consists of the three sites $\{ |g,n\rangle,\ |e,n\rangle,\ |f,n\rangle \}$. The phase $\phi$ acts as an effective magnetic flux, essential for the non-reciprocal coupling. The coupling inhomogeneity originates from the bosonic operators in Eq.~(\ref{ham}). When $2\gamma \gg J_3$, level $|f\rangle$ can be adiabatically eliminated \cite{Adia-elimination}, yielding the effective Hamiltonian,
\begin{equation}
	\label{ham_eff}
	\begin{split}
		\hat{H}_{\text{eff}} = &\, (J_1 + J_2 \hat{a}) \left| e \right\rangle \left\langle g \right| + (J_1 + J_2 \hat{a}^\dagger) \left| g \right\rangle \left\langle e \right| \\
		&- i \frac{J_3^2}{\gamma} \left( e^{i\phi} \left| g \right\rangle \left\langle e \right| + e^{-i\phi} \left| e \right\rangle \left\langle g \right| \right) - i \frac{J_3^2}{\gamma}.
	\end{split}
\end{equation}
The last term is a purely imaginary constant that can be dropped, as it merely contributes an overall amplitude factor without affecting the physical evolution. Therefore, the effective Hamiltonian in the lattice basis $\{ |g,n\rangle,\ |e,n\rangle \}$ becomes,
\begin{equation}
	\label{ham_effnew}
	\begin{split}
		\hat{H}_{\text{eff}} = \sum_{n=0}^{\infty} &\left( J_1 - i \frac{J_3^2}{\gamma} e^{-i\phi} \right) |e,n\rangle\langle g,n| \\
		&+ \sum_{n=0}^{\infty} \left( J_1 - i \frac{J_3^2}{\gamma} e^{i\phi} \right) |g,n\rangle\langle e,n| \\
		&+ \sum_{n=0}^{\infty} J_2 \sqrt{n+1} \left( |e,n\rangle\langle g,n+1| + \text{H.c.} \right).
	\end{split}
\end{equation}
In Eq.~(\ref{ham_effnew}), the first two terms account for nonreciprocal intracell coupling, whereas the third term reflects inhomogeneous intercell coupling, realizing a semi‑infinite variant of the SSH model.

Identifying the non-Hermitian skin effect requires solving the Hamiltonian's eigenproblem. Because a truncated numerical diagonalization of the semi-infinite lattice fails to accurately represent the eigenstates, an analytical approach is necessary. To this end, we outline the analytical solution by first rewriting the effective Hamiltonian in the following form,
\begin{equation}
	\label{ham_effnew2}
	\begin{split}
		\hat{H}_{\text{eff}} = &\, J_2 \left[ (\hat{a}^\dagger - \beta_1) \hat{\sigma}_- + (\hat{a} - \beta_2) \hat{\sigma}_+ \right],
	\end{split}
\end{equation}
with $\beta_{1,2} = -\frac{\alpha_{1,2}}{J_2}$, $\alpha_1 = J_1 - i\frac{J_3^2}{\gamma}e^{i\phi}$, $\alpha_2 = J_1 - i\frac{J_3^2}{\gamma}e^{-i\phi}$, and $\hat{\sigma}_+ = |e\rangle\langle g|$, $\hat{\sigma}_- = |g\rangle\langle e|$ are known as the raising and lowering operators, respectively. The next step is to perform a suitable similarity transformation on Eq.~(\ref{ham_effnew2}). By defining $\hat{S} = e^{-\beta_1 \hat{a} + \beta_2 \hat{a}^\dagger}$, the transformed Hamiltonian reads,
\begin{equation}
	\label{transformation}
	\begin{split}
		\hat{S}^{-1} \hat{H}_{\text{eff}} \hat{S} &= e^{-\beta_2 \hat{a}^\dagger + \beta_1 \hat{a}} \hat{H}_{\text{eff}} e^{-\beta_1 \hat{a} + \beta_2 \hat{a}^\dagger} \\
		&= J_2 (\hat{a}^\dagger \hat{\sigma}_- + \hat{a} \hat{\sigma}_+) = \hat{H}_{\text{JC}}.
	\end{split}
\end{equation}
This Hamiltonian is precisely that of the well-known and exactly solvable Jaynes-Cummings (JC) model \cite{JCmodel}. 

By virtue of the similarity transformation, the eigenvalues of the non-Hermitian effective Hamiltonian $\hat{H}_{\text{eff}}$ are obtained as,
\begin{equation}
	\label{eigenvalue}
	\begin{split}
		E_0 &= 0, \\
		E_{n,\pm} &= \pm J_2 \sqrt{n+1}, \quad (n=0,1,2,\ldots,\infty),
	\end{split}
\end{equation}
with the corresponding right eigenvectors given by, 
\begin{equation}
	\label{righteigenvectors}
	\begin{split}
		|\psi_0^R\rangle &= \hat{S}|0, g\rangle, \\
		|\psi_{n,\pm}^R\rangle &= \frac{1}{\sqrt{2}}\hat{S}(|n, e\rangle \pm |n+1, g\rangle).
	\end{split}
\end{equation}
They satisfy the eigenvalue equations, $\hat{H}_{\text{eff}} |\psi_0^R\rangle = E_0 |\psi_0^R\rangle$ and $\hat{H}_{\text{eff}} |\psi_{n,\pm}^R\rangle = E_{n,\pm} |\psi_{n,\pm}^R\rangle$. 

The complete dynamical evolution further requires the left eigenvectors, which are determined by $\hat{H}_{\text{eff}}^\dagger |\psi_0^L\rangle = E_0^* |\psi_0^L\rangle$ and $\hat{H}_{\text{eff}}^\dagger |\psi_{n,\pm}^L\rangle = E_{n,\pm}^* |\psi_{n,\pm}^L\rangle$, and their explicit expressions are as follows,
\begin{equation}
	\label{lefteigenvectors}
	\begin{split}
		|\psi_0^L\rangle &= (\hat{S}^{-1})^\dagger |0, g\rangle, \\
		|\psi_{n,\pm}^L\rangle &= \frac{1}{\sqrt{2}} (\hat{S}^{-1})^\dagger \left( |n, e\rangle \pm |n+1, g\rangle \right).
	\end{split}
\end{equation}
The left and right eigenvectors form a biorthogonal set, obeying $\langle \psi_0^L | \psi_0^R \rangle = 1$, $\langle \psi_{n,\lambda}^L | \psi_{n,\lambda'}^R \rangle = \delta_{nn',\lambda\lambda'}$, ($\lambda,\lambda'=\pm$). Using the biorthogonal basis, the time evolution from an arbitrary initial state $|\Psi(0)\rangle$ can be described as \cite{Biorthogonal1,Biorthogonal2},
\begin{equation}
	\label{evolution}
	|\Psi(t)\rangle = c_0 |\psi_0^R\rangle + \sum_{n=0}^{\infty} \sum_{\lambda=\pm} c_{n,\lambda} e^{-i E_{n,\lambda} t} |\psi_{n,\lambda}^R\rangle,
\end{equation}
where the superposition coefficients $c_0 = \langle \psi_0^L \vert \Psi(0) \rangle$ and $c_{n,\lambda} = \langle \psi_{n,\lambda}^L \vert \Psi(0) \rangle$.
Thus, via a similarity transformation, both the complete eigensolutions and the dynamical evolution of the model can be obtained in analytic form, thereby avoiding artifacts from numerical truncation. It should be noted that all eigenenergies in Eq.~(\ref{eigenvalue}) are real and appear in positive-negative pairs, reflecting the chiral symmetry of  Eqs.~(\ref{ham_effnew}) and (\ref{ham_effnew2}), i.e., $\hat{\sigma}_{z} \hat{H}_{\text{eff}} \hat{\sigma}_{z} = -\hat{H}_{\text{eff}}$ with $\hat{\sigma}_{z} = | e \rangle \langle e | - | g \rangle \langle g |$.

\section{The confined Skin Effect: Characterization and Dynamics}

\begin{figure}[t]
	\includegraphics{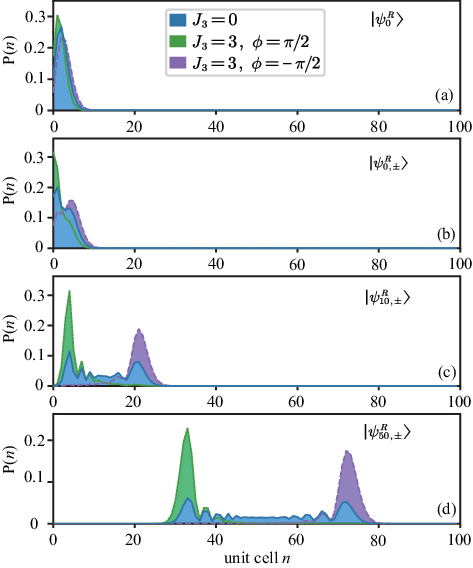}% Here is how to import EPS art
	\caption{\label{figure2}Probability distribution of the normalized right eigenvector with respect to the unit-cell index. Subfigures (a)-(d) respectively display the distributions for the states $|\psi_0^R\rangle$, $|\psi_{0,\pm}^R\rangle$,  $|\psi_{10,\pm}^R\rangle$ and $|\psi_{50,\pm}^R\rangle$, each compared across the three parameter sets: ${J_3=0, \phi=0}$ (blue), ${J_3=3, \phi=\pi/2}$ (green), and ${J_3=3, \phi=-\pi/2}$ (purple). The remaining parameters are fixed at $J_1=1.5$ , $J_2=1$ and $\gamma=50$.}
\end{figure}

\begin{figure}[t]
	\includegraphics{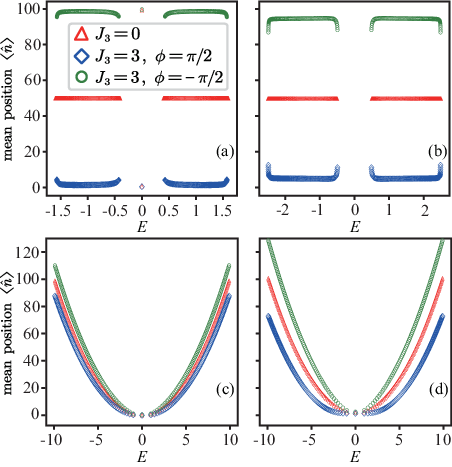}% Here is how to import EPS art
	\caption{\label{figure3}Comparison of the mean position $\langle \hat{n} \rangle$ versus eigenenergy $E$ between uniformly and inhomogeneously coupled nonreciprocal SSH models. Here, $\hat{n} = a^\dagger a$ is the phonon number operator, corresponding to the unit-cell index in the lattice representation. Panels (a,b) present the uniformly coupled model (removing the $\sqrt{n+1}$ coefficient in Eq.~(\ref{ham_effnew}), with total cell number $N = 100$). Panels (c,d) show the semi-infinitely long inhomogeneously coupled model (displaying the zero mode and its nearby 198 modes). Parameters: $J_2 = 1$ and $\gamma=50$ are fixed. Panels (a,c): $J_1 = 0.6$; Panels (b,d): $J_1 = 1.5$. Other settings: $J_3 = 0$ (red open triangles), $J_3 = 3$, $\phi = \pi/2$ (blue open diamonds), $J_3 = 3$, $\phi = -\pi/2$ (green open circles). }
\end{figure}

\begin{figure}[t]
	\includegraphics{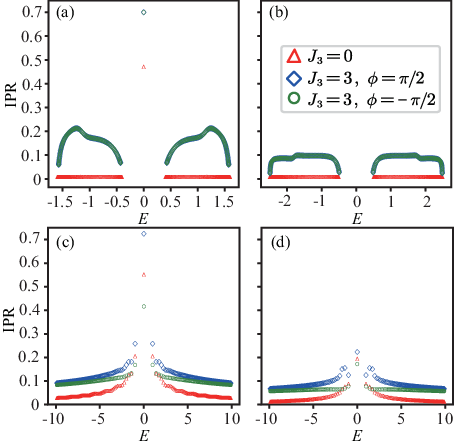}% Here is how to import EPS art
	\caption{\label{figure4}Comparison of the IPR versus eigenenergy $E$ between uniformly and inhomogeneously coupled nonreciprocal SSH models. Panels (a)-(d) use the same parameter sets and share the same marker scheme as the corresponding panels in Fig.~\ref{figure3}.}
\end{figure}

\begin{figure}[t]
	\includegraphics{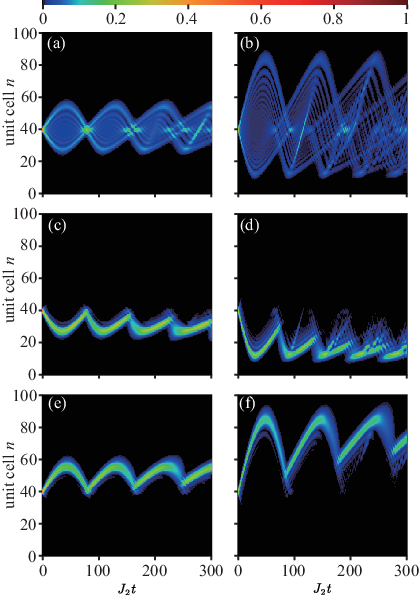}% Here is how to import EPS art
	\caption{\label{figure5}Normalized time evolution in the inhomogeneous SSH model from the initial state $|g,40\rangle$. Parameters: $J_2 = 1$ and $\gamma = 50$ are fixed. The left and right columns correspond to $J_1 = 0.6$ and $J_1 = 1.5$, respectively. From top to bottom, the three rows represent: $J_3 = 0$ (panels (a,b)), $J_3 = 3$, $\phi = \pi/2$ (panels (c,d)), and $J_3 = 3$, $\phi = -\pi/2$ (panels (e,f)).}
\end{figure}

To determine whether the model exhibits the non-Hermitian skin effect, it is essential to analyze the spatial distribution of its eigenstates. This can be conveniently achieved by plotting the analytical expressions of the quantum states. We examine three parameter regimes: $J_3 = 0$, and $J_3 \neq 0$ with $\phi = \pm \pi/2$.

In the first case with $J_3 = 0$, the nonreciprocal term in Eq.~(\ref{ham_effnew}) vanishes, reducing the model to a semi-infinite Hermitian SSH chain \cite{women}. Consequently, the similarity transformation $\hat{S}$ simplifies to a displacement operator $D(\alpha) = e^{\alpha (a^\dagger - a)}$ with $\alpha = -\frac{J_2}{J_1}$. The eigenstates are then obtained by applying this displacement operator to the eigenstates of the Jaynes-Cummings (JC) model. Since the JC eigenstates are either $|g,0\rangle$ or superpositions of $\{|e,n\rangle, |g,n+1\rangle\}$, they are localized within a region spanning at most two unit cells in the Fock-state lattice. While the displacement operator can broaden this distribution in space, the states remain notably more localized compared to the bulk states of a uniform SSH model. This localized character is illustrated by the probability distributions shown in blue in Fig.~\ref{figure2}.

When $J_3 \neq 0$ and $\phi = \pi/2$, the transition from $|e,n\rangle$ to $|g,n\rangle$ dominates over its reverse, causing the probability distribution to shift and compress toward the left. As shown by the green curves in Fig.~\ref{figure2}, this nonreciprocity further enhances the localization originally induced by the inhomogeneous $\sqrt{n}$-dependent coupling. Specifically, it skews the probability distribution so that it is overwhelmingly concentrated on the leftmost side. This behavior differs significantly from that in uniformly coupled models. Here, each eigenmode exhibits skin localization confined to a finite range, unable to reach the physical boundary. The spatial extent of this confinement is directly determined by the profile of the corresponding localized eigenstate for the case of $J_3 = 0$. Similarly, for $\phi = -\pi/2$, a confined skin effect localized to the right emerges, as shown by the purple curves in Fig.~\ref{figure2}.

To better capture the characteristics of the confined non-Hermitian skin effect, we compare it with the conventional skin effect in uniformly coupled systems. The traditional Hermitian, finite-length SSH model exhibits a topological phase ($J_1 < J_2$) with edge states at both ends. Introducing nonreciprocity $J_3$ induces a skin effect that localizes bulk states toward the boundaries, as shown in Figs.~\ref{figure3}(a,b). Notably, in the topological regime, edge states may collapse at an exceptional point (EP) \cite{broken1,explain}. However, in our model with inhomogeneous coupling, the originally localized eigenmodes become further concentrated within a confined spatial range. This range can be tuned by the ratio $J_1/J_2$: a larger ratio enhances the deviation from the original JC-model eigenstates and broadens the extent of localization, as shown in Figs.~\ref{figure3}(c,d). Furthermore, owing to the semi-infinite geometry, only one topological edge state at $E=0$ exists, and it remains robust against variations in $J_1$, $J_2$, and $J_3$.

We further compare the localization of eigenstates under uniform and inhomogeneous couplings, quantified by the inverse participation ratio (IPR) \cite{IPR}. For a normalized quantum state $\vert \psi \rangle = \sum_n \left( a_{g,n}|g,n\rangle + b_{e,n}|e,n\rangle \right)$ with $\sum_n \left( |a_{g,n}|^2 + |b_{e,n}|^2 \right) = 1$ , the IPR is defined as $\text{IPR} = \sum_n \left( |a_{g,n}|^4 + |b_{e,n}|^4 \right)$. Figure~\ref{figure4} shows that the nonreciprocal term indeed enhances the localization. In the uniformly coupled case [Figs.~\ref{figure4}(a,b)], the enhancement is more pronounced for modes in the interior of the band. Under inhomogeneous coupling [Figs.~\ref{figure4}(c,d)], aside from a few states showing rightward skin localization, the enhancement is strongest for states near the zero mode. Overall, for both coupling types, the skin effect becomes more pronounced when $J_3^2 / \gamma$ approaches $J_1$, as verified by comparing panels (a) with (b) and (c) with (d) in Fig.~\ref{figure4}, respectively.

Beyond inducing a confined skin effect on the eigenmodes, inhomogeneous coupling also directly influences the dynamical evolution of quantum states. When $J_3 = 0$, this coupling acts as an effective confining potential, causing a probability distribution initialized at a single site to oscillate only within a finite spatial range. This localization range depends on the ratio $J_1/J_2$: a larger ratio corresponds to more extended eigenmodes and thus a broader oscillation span, as shown in the comparison between Figs.~\ref{figure5}(a) and \ref{figure5}(b). When nonreciprocity is introduced ($J_3 \neq 0$), the eigenmodes exhibit the confined skin effect, becoming skewed to the left ($\phi = \pi/2$) or to the right ($\phi = -\pi/2$). Consequently, the dynamical evolution retains only the trajectory of the probability distribution at the extreme left [Figs.~\ref{figure5}(c,d)] or the extreme right [Figs.~\ref{figure5}(e,f))], respectively. This contrasts with uniformly coupled models, where the entire distribution accumulates fully at the physical boundary \cite{mechanics2}.

\section{Experimental Proposal with a Trapped Ion}

\begin{figure}[t]
	\includegraphics{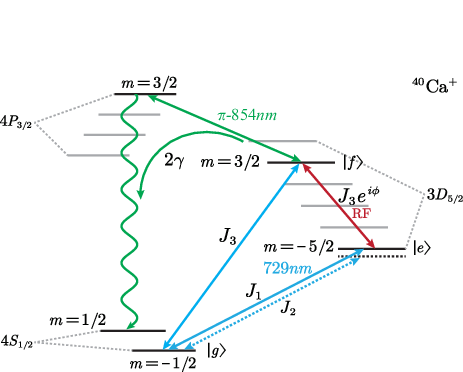}% Here is how to import EPS art
	\caption{\label{figure6}Schematic of the coupled energy levels in a $^{40}\mathrm{Ca}^+$ ion. The states $|g\rangle$, $|e\rangle$, and $|f\rangle$ are selected from specific Zeeman sublevels within the ground and metastable states, forming a triangular coupling scheme. Four coupling processes are involved: three employ frequency-tunable $729\,\mathrm{nm}$ lasers (blue arrows), and one uses a radio-frequency drive (red arrow). Effective decay from the auxiliary level $|f\rangle$ is achieved by coupling it to a short-lived state with an $854\,\mathrm{nm}$ laser (green arrow), followed by spontaneous emission (green wavy arrow) back to a specific ground-state Zeeman sublevel.}
\end{figure}

In this section, we discuss a possible experimental scheme based on a single trapped $^{40}\mathrm{Ca}^{+}$ ion. Trapped ions constitute a prominent platform for quantum simulation owing to their high operational fidelities and long coherence times \cite{simulation1,simulation2,simulation3}. Even a single trapped ion can exhibit a rich internal level structure and highly controllable external motional states. 

Following the model schematic in Fig. 1, the specific level structure of the $^{40}\mathrm{Ca}^{+}$ ion is illustrated in Fig.~\ref{figure6}. The three internal electronic states $|g\rangle$, $|e\rangle$, and $|f\rangle$ are chosen as the levels $|4S_{1/2}, m=-1/2\rangle$, $|3D_{5/2}, m=-5/2\rangle$, and $|3D_{5/2}, m=3/2\rangle$, respectively. The lattice coupling requires four driving fields: $|g\rangle$ and $|e\rangle$ are coupled by two narrow-linewidth $729\,\mathrm{nm}$ lasers, one resonant and the other red‑detuned by the phonon frequency ($\sim 1\,\mathrm{MHz}$). The $|g\rangle$-$|f\rangle$ coupling is also achieved with a resonant $729\,\mathrm{nm}$ laser, while the $|e\rangle$-$|f\rangle$ coupling employs a radio‑frequency electric field \cite{RF}. Moreover, the coupling coefficients $J_1$, $J_2$, $J_3$, and $J_3 e^{i\phi}$ can be tuned to desired values by adjusting the intensity and phase of the driving field. To induce decay from the metastable auxiliary level $|f\rangle$, a $\pi$-polarized $854\,\mathrm{nm}$ laser couples it to the short‑lived $|4P_{3/2}, m=3/2\rangle$ level (lifetime $6.9\,\mathrm{ns}$), which rapidly decays spontaneously into $|4S_{1/2}, m=1/2\rangle$. This process yields an effective decay rate of $2\gamma$.

The original Hamiltonian describing the coupling between the radiation field and the internal and external states of the ion is rather complex. To derive the simplified form of Eq.~(\ref{ham}), the system must operate within the Lamb-Dicke regime \cite{trappedion}, where $\eta^2(2n + 1) \ll 1$. Here, the Lamb-Dicke parameter is defined as $\eta = k_L\sqrt{\hbar/(2\nu M)}$, with $M$ being the ion's mass, $\nu$ the ion's vibrational frequency, and $k_L$ the wave vector of the incident field. For $\eta \sim 0.05$, up to $40$ effective unit cells become available. If the initial state is chosen as $|g,10\rangle$, with coupling coefficients set to $J_1 = 0.6$, $J_2 = 1$, and $J_3 = 3$, the probability distribution of the dynamical evolution remains within $25$ phonons. While the platform of $40$ unit cells can theoretically exhibit either the confined skin effects toward the left or toward right, current experimental capabilities allow precise control and measurement of phonon numbers up to about $30$ \cite{fockdistribution}, which exceeds the required range and is thus sufficient to capture the full evolution. Therefore, both leftward and rightward skin effects are experimentally observable. Note that the effective Hamiltonian Eq.~(\ref{ham}) holds only in the absence of decay. This requires post-selection to discard experimental outcomes in which spontaneous emission fluorescence is detected.

\section{Conclusion}
In this work, we investigate the influence of inhomogeneous coupling on the non-Hermitian skin effect in a semi-infinite Fock-state lattice. Our results show that such inhomogeneity localizes the bulk states of an otherwise uniform SSH model, thereby confining the skin effect to a finite spatial region. An explicit analytical solution fully reveals this behavior, and we systematically analyze the spatial profiles of the localized skin modes and dynamical evolution of the corresponding confined non-Hermitian skin effect. A feasible experimental scheme based on an single trapped ion is also proposed. This study deepens the understanding of topological transport in inhomogeneous non-Hermitian systems and provides a new approach for realizing tunable skin effects in quantum simulators.

\begin{acknowledgments}
Z. J. Deng is grateful to useful discussion with Xin-Sheng Chen. This work was supported by the National Natural Science Foundation of China under Grants No. 12174448, No. 12574403 and No. 11574398.
\end{acknowledgments}

% whether or not they are actually referenced in the text. This is ap

%

\end{document}